\documentclass[apjl,a4paper,12pt,useAMS]{emulateapj}
\usepackage{txfonts}
\usepackage{amssymb}

\usepackage{graphicx,graphics}
\usepackage{natbib}

\def\cm{\,\rm cm}
\def\s{\,{\rm s}}

\def\erg{\,{\rm erg}}

\begin{document}
\title{The star formation history inferred from long gamma-ray bursts with high pseudo-redshifts}


\author{Wei-Wei Tan$^{1}$,  Xiao-Feng Cao$^{2}$, Yun-Wei Yu$^3$}

\address{$^1$ School of Astronomy and Space Science, Nanjing University,
Nanjing 210093, China\\Email: wwtan@nju.edu.cn\\
$^2$ School of Physics and Electronics Information, Hubei University of Education, 430205, Wuhan, China\\
$^3$ Institute of Astrophysics, Central China Normal University,
Wuhan 430079, China;\\Email: yuyw@mail.ccnu.edu.cn}

\begin{abstract}
By employing a simple semi-analytical star formation model where the
formation rates of Population (Pop) I/II and III stars can be
calculated, respectively, we account for the number distribution of
gamma-ray bursts (GRBs) with high pseudo-redshifts that was derived
from an empirical luminosity-indictor relationship. It is suggested
that a considerable number of Pop III GRBs could exist in the
present sample of {\it Swift} GRBs. By further combining the
implication for the star formation history from the optical depth of
the CMB photons, it is also suggested that only a very small
fraction ($\sim 0.6\%$) of Pop III GRBs could have triggered the
{\it Swift} BAT. These results could provide an useful basis for
estimating future detectability of Pop III stars and their produced
transient phenomena.

\end{abstract}





\section{Introduction}

Gamma-ray bursts (GRBs), especially the major class of a duration
longer than 2 seconds, are the most luminous objects in the
universe, which makes them detectable even at the edge of the
universe. On one hand, the reported highest redshift of GRBs so far
is up to $z\sim 9.4$ for GRB 090429B (Cucchiara et al. 2011). On the
other hand, some theoretical models predict that GRBs could be
detected at much more distance (Band 1993; Bromm \& Loeb 2006; de
Souza et al. 2011). Therefore, it is widely suggested that long GRBs
can be used as a cosmological tool to probe the early universe.

Long GRBs are also known to be associated with the collapse of
massive stars (Galama et al. 1998; Stanek et al. 2003; Hjorth et al.
2003), which indicates that the GRB event rates could trace the
cosmic star formation history either unbiasedly (e.g., Chary et al.
2007) or, more probably, with an additional evolution effect (Daigne
et al. 2006; Ksitler 2008; Salvaterra 2009; Campisi et al 2010).
Thanks to the launch of the {\it Swift} spacecraft, the number of
GRBs with a measured redshift has been increasing rapidly during the
past decade. This makes it possible to clarify the connection
between the GRB numbers and the star formation rates (SFRs; e.g.,
Kistler et al. 2008; Cao et al. 2011; Wang \& Dai 2011; Tan et al.
2013) and finally to infer the SFRs at high redshifts (Totani 1997;
Wijers et al. 1998; Lamb \& Reichart 2000; Porciani \& Madau 2001;
Murakami \& Yonetoku 2005; Y\"{u}ksel et al. 2008; Kistler et al.
2009; Wang \& Dai 2009; Ishida et al. 2011; Wang et al. 2013) where
a direct SFR-measurement is extremely difficult. Nevertheless, such
attempts could somewhat be disturbed/hindered by the limited number
of measured redshifts and, in particular, by some inevitable
observational selection effects (Guetta \& Piran 2007; Cao et al.
2011).

As an alternative and complementary method, in Tan et al. (2013), we
proposed to estimate pseudo-redshifts for the {\it Swift} GRBs with
an empirical relationship between the spectral peak energy and the
peak luminosity. Consequently, 498 pseudo-redshifts up to $z\sim30$
were derived from the Butler's GRB catalog where the spectral peak
energies are roughly estimated with Bayesian statistics rather than
observed. Although the $L-E_p$ relationship is not so tight and the
obtained pseudo-redshifts are not so reliable individually, a
statistical study on the pseudo-luminosity distributions of the GRBs
can in principle be implemented, especially worthy of mention, for
different narrow redshift ranges. As a result, the luminosity
function of the GRBs was found to be strongly redshift-dependent,
for redshifts $z\leqslant3.5$ and a previously-determined star
formation history. From the pseudo-redshift sample of Tan et al.
(2013), we can find that there are 38 GRBs whose pseudo-redshifts
are higher than $10$, where 23 GRBs are above the lower luminosity
cut off of {\it Swift} satellite. This statistically indicates that
a remarkable number of high-redshift GRBs may exist in the present
{\it Swift} GRB sample, most of which\textbf, however, evaded direct
redshift measurements due to redshift selection effects.

As widely considered, for redshifts $z\gtrsim10$, high mass
Population (Pop) III stars are probably more dominant than normal
Pop I/II stars (Bromm et al. 2002; Abel et al. 2002). Pop III stars
were born in metal free gas and their deaths started the metal
enrichment of the intergalactic medium (IGM) via supernova feedback,
which subsequently lead to the formation of Pop I/II stars (Ostriker
\& Gnedin 1996; Madau et al. 2001; Bromm et al. 2003; Furlanetto \&
Loeb 2003). The high mass and zero-metallicity of Pop III stars make
them easily to produce a collapsar GRB (Hirschi 2007), whose
isotropically-equivalent energy could be as high as $\sim
10^{55-57}\rm ergs$ (\rm M\'esz\'aros \& Rees 2010; Suwa \& Ioka
2011). Additionally, as the first generation stars in the universe,
Pop III stars also turn on the cosmic reionization by emitting
ultraviolet photons, and so the reionization history should strongly
depend on the formation history of Pop III stars (Furlanetto \& Loeb
2005; Barkana 2006; Robertson et al. 2010).

Therefore, in this paper we try to use the GRBs with high
pseudo-redshifts ($z> 3.5$) to probe the SFRs up to the beginning of
the reionization era as well as the detection efficiency of Pop III
GRBs, by combining the constraint from the optical depth of the
cosmic microwave background (CMB) photons. In the next section, we
will estimate the high-redshift SFRs by a usual method where the
contribution to GRBs by Pop III stars is ignored, which could lead
to an unacceptable result. In contrast, in section 3 we attempt to
understand the GRB numbers by considering the GRB productions of
both Pop I/II and III stars, where a simple semi-analytical model
for the collapse of dark matter halos is employed. Finally, the
results are given and discussed in section 4.

\begin{table}
\centering \caption{The GRB numbers in different redshift ranges.}
\begin{tabular}{c|cc}
\hline \hline
Redshift & $N_{\rm obs}^\dag$&$N_{\rm p}^\ddag$\\
\hline
$0\sim0.5$&10~ & 28\\%
$0.5\sim1$&30~ & 74\\%
$1\sim1.5$&22~ & 96\\%
$1.5\sim2$&23~ & 65 \\%
$2\sim2.5$&20~ & 37\\%
$2.5\sim3$&19~ & 37 \\%
$3\sim3.5$&14~ & 24\\%
$3.5\sim4.5$&12~ & 24 \\%
$4.5\sim5.5$&5~ & 14\\%
$5.5\sim6.5$&2~ & 7 \\%
$6.5\sim8$&2~ & 9 \\%
$8\sim10$&1~ & 11\\%
$10\sim14$&0~ & 12\\%
$14\sim18$&0~ & 5\\%
$18\sim22$&0~ & 2\\%
$22\sim26$&0~ & 1 \\%
$26\sim30$&0~ & 3 \\%
\hline\hline\\
\end{tabular}\\
$^\dag$GRBs of observed redshift.\\
$^\ddag$GRBs of a pseudo-redshift.\label{T1}
\end{table}
\section{SFR-determination without Pop III stars}
Table 1 lists the numbers of GRBs with observational or
pseudo-redshifts for different redshift ranges, which are taken from
Tan et al. (2013). Here only GRBs with luminosities higher than the
lower luminosity cut off of {\it Swift} are taken into account. The
difference between these two sets of numbers arises from the
selection effects of redshift measurements that are\textbf, however,
difficult to be described theoretically. In this paper, we will only
pay attention to the numbers of pseudo-redshifts in order to probe
the high-redshift ranges and avoid the redshift selection effects.

\begin{figure}
\centering\resizebox{0.4\textwidth}{!}{\includegraphics{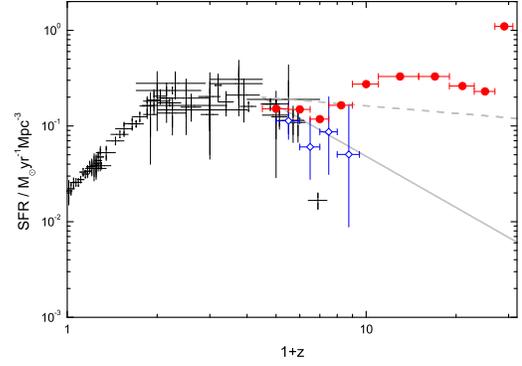}}
\caption{SFRs inferred from the GRBs with pseudo-redshifts $>3.5$
(solid circles), where the contribution to the GRBs by Pop III stars
is ignored. The crosses display the SFRs according to Hopkins \&
Beacom (2006) and the open diamonds correspond to the GRB-inferred
SFRs given by Kistler et al. (2009). The solid (or dashed) line
represents a power-law star formation history, which is required by
the CMB optical depth for $N_{\gamma}=4000$ and $f_{\rm esc}=10\%$
(or $f_{\rm esc}=2\%$).}\label{L-Ep}
\end{figure}

In view of the model, for a given GRB luminosity function
$\Phi_z(L)$ and GRB event rate $\dot{R}(z)$, the expected number of
GRBs within redshift range $z_1<z< z_2$ can be calculated by
\begin{eqnarray}
N_{[z_1, z_2]}&=&{\Delta\Omega\over 4\pi} T
\int^{z_2}_{z_1}\int_{L_{\rm lc}(z)}\Phi_z(L)\dot{R}(z)dL{dV(z)\over
1+z}, \label{expected number}
\end{eqnarray}
where $\Delta\Omega/4\pi\sim 0.1$ is the field view of the Burst
Alert Telescope (BAT) on board {\it Swift}, $T\sim7 \rm yr$ is the
observational period, $L_{\rm lc}(z)$ is an adopted lower cutoff of
the GRB luminosity corresponding to the selected data [see Equation
(3) in Tan et al. (2013) and explanation therein], and $dV(z)$ is
the comoving cosmic volume.

From the GRBs having a pseudo-redshift $z\leqslant3.5$ and the
corresponding star formation history
\begin{equation}
\dot{\rho}_{*}(z)\propto\left\{
\begin{array}{ll}
(1+z)^{3.44},~z\leqslant 0.97,\\
(1+z)^0,~~~~~0.97<z\leqslant 3.5,\\
\end{array}\right.
\end{equation}
Tan et al. (2013) derived a GRB luminosity function by:
\begin{eqnarray}
\Phi_z(L)={1\over 4.5 L_{\rm
b}(z)}\left\{~\begin{array}{ll}\left({L\over
L_b(z)}\right)^{-0.8},~~~~& L\leqslant L_b(z),\,\\
\left({L\over
L_b(z)}\right)^{-2},~~~~&L>L_{b}(z),\,\end{array}\right. \label{LF}
\end{eqnarray}
where the break luminosity reads $L_{\rm
b}(z)=1.2\times10^{51}(1+z)^2\erg \s^{-1}$, and a GRB event rate as
\begin{eqnarray}
\dot{R}(z)=f_Bp(z)\dot{\rho}_*(z),\label{GRB rate}
\end{eqnarray}
where the proportionality coefficient $f_Bp(z)=2.4\times10^{-8}
(1+z)^{-1}M_{\odot}^{-1}$ with $f_B$ being the beaming factor of the
GRB jets and $p(z)$ representing the GRB-production efficiency of
the stars.

As the most straightforward assumption in this paper, we assume that
Equations (3) and (4) are valid for all redshifts, which could be
correct if all GRB progenitors belong to the same type of stars
(i.e., Pop I/II stars). Then following Kistler et al. (2008, 2009),
the SFRs at high redshifts can be estimated by the following
equation
\begin{eqnarray}
\dot{\rho}_{*}(z)\approx{ N_{[z-{\Delta z\over2},z+{\Delta
z\over2}]}\int^{3.5}_{0}\int_{L_{\rm
lc}(z)}\Phi_z(L)p(z)\dot{\rho}_{*}(z)dL{dV(z)\over 1+z}\over
N_{[0,3.5]}\int_{z-{\Delta z\over2}}^{z+{\Delta
z\over2}}\int_{L_{\rm lc}(z)}\Phi_z(L)p(z)dL{dV(z)\over 1+z}}.
\end{eqnarray}
Substituting the numbers of GRBs with pseudo-redshifts for different
redshift ranges into the above equation, we derive the high-redshift
($z>3.5$) SFRs as shown by the solid circles in Figure 1, where a
big bump appears surprisingly in the cosmic star formation history
within the redshift range $8\lesssim z\lesssim 30$.

By considering the connection between the star formation and cosmic
reionization, the SFRs can also be constrained by the CMB optical
depth $\tau=0.088\pm 0.015$ measured by the Wilkinson Microwave
Anisotropy Probe (WMAP) experiment. Supposing a simple power-law
high-redshift star formation history as $\dot{\rho}_{*}(z)\propto
(1+z)^{-\alpha}$ for $z> 3.5$, Yu et al. (2012) derived a constraint
on the index $\alpha$ as
\begin{eqnarray}
\alpha=2.18\lg (N_{\gamma}f_{\rm esc})-3.89,
\end{eqnarray}
where $N_{\gamma}$ is the number of ionizing ultraviolet photons
released per baryon of the stars and $f_{\rm esc}$ is the fraction
of these photons escaping from the stars. For a typical value of
$N_{\gamma}\sim4000$ for a Salpeter stellar initial mass function
and a metallicity $0.05Z_{\odot}$ (e.g. Barkana 2001), we can get
$\alpha=1.78$ and $0.26$, corresponding to a typical escaping
fraction $f_{\rm esc}\sim 10\%$ and a very small one $f_{\rm
esc}\sim 2\%$, respectively. Such requirements on the SFRs are shown
by the lines in Figure 1, which are, however, significantly lower
than the GRB-inferred SFRs. In other words, the SFRs simply inferred
from the GRB numbers would lead to an overionized universe, unless
the photon escaping fraction is unacceptably small\footnote{For an
escaping fraction less than 2\%, the CMB optical depth would require
much higher SFRs ($\alpha<0.26$). However, in such cases, the cosmic
reionization can only be complete at redshifts $<6$, which is
inconsistent with the result from the probes to the Gunn-Peterson
trough (Ly-$\alpha$ absorption) toward high-redshift quasars and
galaxies (Yu et al. 2012).}. This indicates that something wrong
must appear in the above processes of determining the SFRs by GRBs.
The most probable reason could be that it is misleading to assume
Equations (3) and (4) to be valid for all GRBs. In fact, for
redshifts $\gtrsim10$, Pop III stars could play a dominant role in
producing GRBs. Due to the unique properties of Pop III stars, the
formation history of Pop III GRBs is probably very different from
that of Pop I/II GRBs. In such a case, one may worry whether the Pop
III GRBs can obey the same $L-E_p$ relationship as Pop I/II GRBs. In
our opinion, the $L-E_p$ relationship is probably determined by the
emission mechanism of the GRB jets, which could not be sensitive to
the progenitor properties.

\section{Confronting the GRB numbers with a star formation model}
\subsection{Semi-analytical star formation model}
In the hierarchical formation model, star formation takes place
during the collapsing and merging of dark matter halos. A
straightforward semi-analytical approach for the abundance of dark
halos was first given by the Press-Schechter (PS) formalism (Press
\& Schechter 1974), which was subsequently improved by Sheth \&
Tormen (1999). By using the Sheth-Tormen mass function (ST) $n_{\rm
ST}(M,z)$ of dark halos, the collapse fraction of mass available for
star formation can be calculated by
\begin{eqnarray}
f_{\rm col}(z)=\int_{M_{\min}}^{\infty}n_{\rm ST}(M,z) {M\over
\bar{\rho}} dM ,\label{fraction}
\end{eqnarray}
where $\bar{\rho}$ is the mean density of the universe and
$M_{\min}$ is the minimum mass of the halos below which the halos
can not collapse. By requiring the virial temperature $T_{\rm vir}$
of the halos to be higher than the temperature permitted by possible
cooling channels, the minimum halo mass can be determined to
$M_{\min}\sim10^8 {\rm M_{\odot}}\left({T_{\rm vir}/ 10^4 \rm
K}\right)^{3/2}\left[{(1+z)/10}\right]^{-3/2}$. Here we typically
assume the minimum halo mass corresponds to the virial temperature
of $T_{\rm vir}=10^4$ both for Pop I/II and Pop III stars, above
which atomic hydrogen cooling is effective.

The star forming halos (``galaxies'') could launch a wind of
metal-enriched gas with an initial speed of $v_{\rm 0}$ by producing
a large number of supernovae. The well-known Sedov (1959) solution
predicts that the galactic wind could travel a comoving distance of
$R_{\rm w}=v_{\rm w}t=(3E_{\rm w}/2\pi f_b {\rho})^{1/5}t^{2/5}$,
where $v_{\rm w}$ is the average wind speed while traveling in the
apace, $\rho$ is the density, $E_{\rm w}={1\over2} f_* f_{\rm w}f_bM
v_{\rm 0}^2$ is the total energy of the galactic wind with
$f_b=0.167$ being the mass fraction of baryonic matter,
 $f_*$ the star formation efficiency, and $f_{\rm w}$ the energy fraction
that goes into the wind. Here we have assumed that each wind has
propagated for half of the age of the universe with ${\rm
t}\approx1/3H(z)$. Therefore, the volume-filling fraction of the
metal-enriched gas of a collapsing halo can be expressed by
$\eta(M,z)=(R_{\rm w}/R_h)^3$, where
$R_h=(3M/720\pi\bar{\rho})^{1/3}$ is the radius of the whole halo
(e.g. Johnson 2010). Then, by considering the mass distribution of
the halos, the total fraction of the cosmic volume that is
metal-enriched by the galactic winds can be written as (Furlanetto
\& Loeb 2005; Greif \& Bromm 2006) \begin{eqnarray}
Q_e(z)=[1+\xi_{\rm hh}(z)]\int_{M_{\min}}^{\infty} \eta(M,z) n_{\rm
ST}(M,z) {M\over \bar{\rho}} dM,
\end{eqnarray}
where $\xi_{\rm hh}(z)$ is the linear correlation function between
two halos with mass M at a comoving distance R:
\begin{eqnarray}
\xi_{\rm hh}(M,R,z)=b^2(M,z)\xi_{\rm mm}(R),
\end{eqnarray}
here $b(M,z)$is the linear bias of a halo and $\xi_{\rm mm}(R)$ is
the mass correlation function (e.g. Mo $\&$ White 2002) with $R$
being the average wind size in the comoving unites. The average bias
of the enriched regions is given by,
\begin{eqnarray}
\bar{b}(z)={\int^{\infty}_{M_{\rm min}}   {\rm d}M M \eta(M,z)
b(M,z) n_{\rm ST} \over \int^{\infty}_{M_{\rm min}} {\rm d}M M
n_{\rm ST}}.
\end{eqnarray}
By assuming that the host galaxies were randomly distributed, we
could get the enrichment probability $p_e=1-{\rm exp}[-Q_e(z)]$.

 Since the halos hosting zero-metallicity
Pop III stars should not be located in the wind radius of old
galaxies, the SFR of Pop III stars at redshift $z$ can be roughly
estimated by
\begin{eqnarray}
\dot{\rho}_{*}^{\rm III}(z)=f_{*}^{\rm
III}f_b\bar{\rho}(1-p_e){{df_{\rm col}}\over dz}{dz\over
dt},\label{SFR3}
\end{eqnarray}
where $f_{*}^{\rm III}$ is the fraction of baryonic mass that goes
into Pop III stars. Simultaneously, the SFR of Pop I/II stars read
\begin{eqnarray}
\dot{\rho}_{*}^{\rm I/II}(z)= f_{*}^{\rm I/II}f_b\bar{\rho}p_e
{{df_{\rm col}}\over dz}{dz\over dt},\label{SFR2}
\end{eqnarray}
where $f_{*}^{\rm I/II}$ is the baryonic fraction for Pop I/II star
formation.

\begin{figure}
\centering\resizebox{0.4\textwidth}{!}{\includegraphics{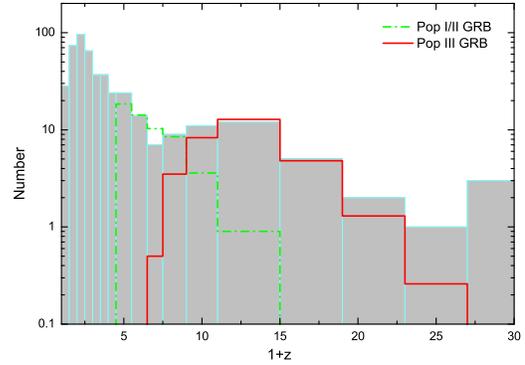}}
\caption{Fitting to the number distribution of the GRBs with
pseudo-redshifts (grey histograms) by combining the contributions to
GRB productions from Pop I/II (dash-dotted line) and III (solid
line) stars.}\label{n-n}
\end{figure}

\subsection{Fitting to the GRB number distribution}
Different from the treatment in Section 2, here we assume the GRB
luminosity function presented in Eq. (3) and the connection between
the GRB event rate and SFR in Eq. (4) are only valid for the GRBs
produced by Pop I/II stars. Instead of Eq. (1), we calculate the
number of Pop III GRBs within redshift range $z_1< z< z_2$ by the
following formula:
\begin{eqnarray}
N^{\rm III}_{[z_1, z_2]}&=&{\Delta\Omega\over 4\pi} T \zeta
\int^{z_2}_{z_1} \dot{R}^{\rm III}(z){dV(z)\over 1+z}
dz\label{numberIII}
\end{eqnarray}
with $\dot{R}^{\rm III}(z)= f_Bp^{\rm III}{\dot{\rho}_{*}^{\rm
III}(z)}$, where the parameter $\zeta$ represents the detection
efficiency of Pop III GRBs (see discussions at the end of Section
3.3) and the GRB-production efficiency of Pop III stars $p^{\rm
III}$ is considered to mainly arise from the mass requirement on the
progenitors. The masses of Pop III stars are usually considered to
be higher than $100\rm M_{\odot}$ (Bromm et al. 2002; Abel et al.
2002), but a much lower mass as $\sim 10\rm M_{\odot}$ was also
suggested by some recent more elaborate researches where the
feedback effects are taken into account simultaneously (Greif \&
Bromm 2006; de Souza 2011; Hosokawa et al. 2011, 2012; Stacy et al.
2012). In such a case, there could be only a small fraction of Pop
III stars available for GRB production, because the GRB production
could require the progenitor's mass higher than $\sim 40\rm
M_{\odot}$ (Belczynski et al. 2007; Tornatore et al. 2007; Marassi
2009). In addition, it is widely suggested that Pop III stars with
masses $140 \rm M_{\odot}\lesssim m \lesssim 260M_{\odot}$ would be
completely disrupted by a supernova explosion due to the
pair-creation instability (PISN; Heger \& Woosley 2002), so they
also can not contribute to GRBs. Finally, by invoking the Salpeter
mass function $\phi(m)$, the GRB production efficiency of Pop III
stars can be estimated by $p^{\rm III}=\left.\left[\int_{40
M_\odot}^{140 M_\odot}\phi(m)dm+\int_{140M_\odot}^{10^3
M_\odot}\phi(m)dm\right]\right/\int_{10 M_\odot}^{10^3
M_\odot}M\phi(m) dm=4\times10^{-3}\rm M_\odot^{-1}$.

Substituting Eq. (12) into (1) and Eq. (11) into (13), we can
calculate the numbers of Pop I/II and III GRBs, respectively, for
specific values of model parameters. By confronting these
model-predicted GRB numbers with the numbers listed in Table 1, we
can in principle obtain an observational constraint on the model
parameters. We firstly take typical values for some parameters
according to some previous works as $T_{\rm vir}=10^{4}$ K for
atomic hydrogen cooling (Haiman et al. 2000; Furlanetto \& Loeb
2005), $f_{\rm w}=0.1$ (e.g., Scannapieco \& Broadhursts 2001;
Springel \& Hernquist 2003; ), and $v_{ 0}=200\rm km/s$ (Fabian et
al. 1980; Eymeren et al. 2007). For the remaining more uncertain
parameters $f_{*}^{\rm I/II}$, $f_{*}^{\rm III }$, and $\zeta$,
their best-fit values are constrained to $f_{*}^{\rm I/II}=0.16$ and
$f_{*}^{\rm III}\zeta=6\times10^{-6}$ by fitting to the GRB number
distribution without error bars. The corresponding fitting is shown
in Figure 2 by the lines. The value of $f_{*}^{\rm I/II}$ could be
well consistent with some previous analytical estimations and
simulations (Greif \& Bromm 2006; Marassi et al. 2009).
Qualitatively, our result indicates that (i) the formation history
of Pop I/II stars with typical values of parameters is basically
favored by the GRB numbers, and (2) a considerable number of Pop III
GRBs could exist in the present {\it Swift} GRBs although the
degeneracy between $\zeta$ and $f_{*}^{\rm III}$ has not been
removed.

\begin{figure}
\centering\resizebox{0.4\textwidth}{!}{\includegraphics{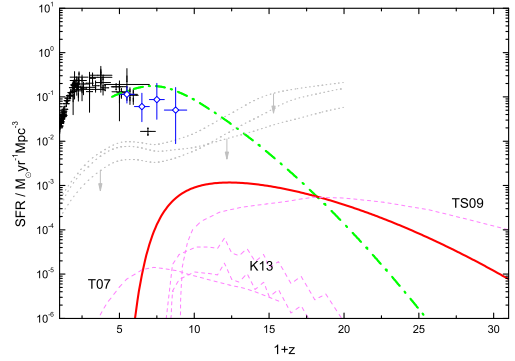}}
\caption{High-redshift star formation history of Pop I/II
(dash-dotted line) and III stars (solid line), which are constrained
by the GRB numbers and the CMB optical depth together. For
comparisons, some simulation results of the formation history of Pop
III stars are also presented by dashed lines, such as Tornatore et
al. (2007; T07), Trenti \& Stiavelli (2009; TS09), and Kulkarni et
al. (2013; K13). The dotted lines with arrows give an upper limit on
the SFRs of Pop III stars, which is taken from Inoue et al.
(2013).}\label{n-n}
\end{figure}

\subsection{Cosmic reionization}
In order to remove the degeneracy between the parameters $\zeta$ and
$f_{*}^{\rm III}$, we further invoke an implication for the star
formation history from the CMB optical depth. The evolution of the
cosmic reionization denoted by $x\equiv n_{\rm H~II}/n_{\rm H}$ can
be determined by the following equation
\begin{eqnarray}
{dx\over dz}=\left[{\dot{n}_{\gamma} \over (1+y)n_{\rm
H}^0}-\alpha_{B}C(z) (1+z)^3(1+y)n_{\rm H}^0 x\right]{dt\over dz}
\end{eqnarray}
with a rate of ionizing ultraviolet photons escaping from stars into
IGM $\dot{n}_{\gamma}(z)=N_{\gamma}f_{\rm esc}{\dot{\rho}_*(z)/
m_B}$, where $m_B$ is the mass of a baryon. $y=0.08$ by assuming
that the helium was only once ionized, $n_{\rm H}^0=1.9\times
10^{-7}~\cm^{-3}$ is the local number density of hydrogen,
$\alpha_{B}=2.6\times10^{-13}\cm^3\s^{-1}$ is the recombination
coefficient for electron temperature of about $10^4K$, and
$C(z)=2.9\left[{(1+z)/ 6}\right]^{-1.1}$ (or $=2.9$) for $z>5$ (or
$z\leqslant5$) is the clumping factor of ionized gas. With a given
reionization history $x(z)$, the CMB optical depth can be calculated
by integrating the electron density times the Thomson cross section
along proper length as
\begin{eqnarray}
\tau=-(1+y)\sigma_{T}n_{\rm H}^0 c\int_0^{z_h}(1+z)^3 x(z){dt\over
dz}dz.
\end{eqnarray}
Here we define the upper limit of the integral to be a moderate
value with $z_h\sim 30$, because the CMB optical depth could be
mainly contributed by the electrons at relatively low redshifts
$z\ll z_h$, and also as the WMAP experiment is insensitive to too
high redshifts (Larson et al. 2011).

In the star formation model with the parameters constrained above,
the remaining free parameter $f_{*}^{\rm III}$ can be determined to
be $f_{*}^{\rm III}=10^{-3}$ from the CMB optical depth
$\tau=0.088$, where we take $f^{\rm I/II}_{\rm esc}=6\%$,
$N_{\gamma}^{\rm I/II}=4000$, $f^{\rm III}_{\rm esc}=30\%$, and
$N_{\gamma}^{\rm III}=3\times 10^4$ (Greif \& Bromm 2006). Such a
value is just located within its theoretically-expected range of
$\sim 10^{-6}-10^{-3}$ (Greif \& Bromm 2006; Marassi et al. 2009).
At the same time, the detection efficiency of Pop III GRBs is
revealed to be $\zeta=0.6\%$. Such a low efficiency could be caused
by various reasons. For example, the luminosity selection with an
unknown luminosity function of Pop III GRBs should be included in
the parameter $\zeta$, and some Pop III GRBs might occur in
extremely dense accretion envelopes that suppress the GRB luminosity
at early times. Moreover, the high gamma-ray variability of the GRBs
could be significantly smoothed by the high redshifts, which could
lead to facilities such as {\it Swift} BAT not to be triggered by
the gamma-ray emission.

\section{Results and Discussions}
The star formation histories of Pop I/II and III stars with the
obtained parameter values are presented in Figure 3. Firstly, the
SFRs at redshift range $5\lesssim z \lesssim10$ are suggested to be
slightly higher than the ones given by Kistler et al. (2009),
because here an evolving GRB luminosity function is adopted for Pop
I/II GRBs and a larger GRB number with pseudo redshift is used here.
Secondly, the formation history of Pop III stars is demonstrated to
be basically consistent with some previous simulations and an
observational upper limit is derived from the gamma-ray
attenuations. To be specific, the peak of the SFRs of Pop III stars
appears from our analysis to be at $z\sim11$ and the formation of
Pop III stars could continue down to as low as $z\sim 5$. The
transition from the formation of Pop III stars to Pop I/II stars
takes place at $z\sim18$. Finally, although the simple star
formation model can account for the GRB numbers for a wide redshift
range as $z\lesssim25$, the GRB number for $26<z<30$ still
significantly exceeds the model prediction. Such an obvious excess
may indicate some other GRB origins in the early universe, e.g.,
superconducting cosmic strings (Cheng et al. 2010). In this paper,
we do not provide a precise regression analysis of the data, by
considering that the large intrinsic uncertainty of the present
pseudo-redshift data makes it unnecessary to find a high-precision
fitting. In the future, a more accurate method to get the GRB
pseudo-redshift is expected.

The determination of the SFRs of Pop III stars is of fundamental
importance for estimating the detection efficiencies of these stars
and their produced transient phenomena with some future facilities.
For example, the upcoming {\it James Webb Space Telescope (JWST)} is
expected to be able to detect the supernovae of Pop III stars up to
$z\sim50$, especially the PISN (Mackey et al. 2003; Weinmann \&
Lilly 2005). For the Pop III GRBs, their radio afterglows may also
be observable by SKA, EVLA, LOFAR, and ALMA in the future (de Souza
et al. 2011). Such an afterglow observation could provide a more
direct test on the gamma-ray detection efficiency ($\zeta$) of the
GRBs.

\section*{Acknowledgments}
The authors thank F. Y. Wang for useful discussions. This work is
supported by the National Basic Research Program of China (973
program, No. 2014CB845800), the National Natural Science Foundation
of China (Grant Nos. 11103004 and 11303010), the Founding for the
Authors of National Excellent Doctoral Dissertations of China (Grant
No. 201225), and the Program for New Century Excellent Talents in
University (Grant No. NCET-13-0822).


\begin{thebibliography}{99}

\bibitem{Abel 2002}Abel, T., Bryan, G. L., \& Norman, M. L. 2002,
Science, 295, 93

\bibitem{Band 93} Band, D., Matteson, J., Ford, L. et al. 1993, ApJ, 413, 281

\bibitem{Barkana 01}Barkana, R., \& Loeb, A. 2001, Phys. Rep., 349, 125

\bibitem{Barkana 06}Barkana, R. 2006, Science, 313, 931

\bibitem{Belczynski 07}Belczynski, K., Bulik, T., Heger, A. and
Fryer, C. 2007, ApJ, 664, 986


\bibitem{Bromm 02}Bromm, V., Coppi, P. S., Larson, R. B. 2002, ApJ,
564, 23

\bibitem{Bromm 03}Bromm, V., Loeb, A. 2003, Nature, 425, 812

\bibitem{Bromm 06}Bromm, V., Loeb, A. 2006, ApJ, 642, 382

\bibitem{Campisi 07}Campisi, M. A., Li, L.-X., Jakobsson, P. 2010,
MNRAS, 407, 1972

\bibitem{Cao 11}Cao, X. F., Yu, Y. W., Cheng, K. S., \& Zheng, X. P. 2011, MNRAS, 416, 2174

\bibitem{Chary 07}Chary, R.-R., Berger, E., \& Cowie, L. 2007, ApJ, 671, 272

\bibitem{Cheng 10} Cheng, K. S., Yu, Y. W., \&
Harko, T. 2010, PRL, 104, 241102

\bibitem{Cucchiara 11}Cucchiara, A., Levan, A. J., Fox, D. B., et
al. 2011, ApJ, 736, 7


\bibitem{Daigne 06}Daigne, F., Tossi, E. M., Mochkovitch, R. 2006,
MNRAS, 372, 1034

\bibitem{de Souza 11}de Souza, R. S., Yoshida, N., Ioka, K. 2011, A\&A,
533, 32

\bibitem{Fabian 80}Fabian, A. C., Nulsen, P. E. J., \& Stewart, G.
C. 1980, Nature, 287, 16

\bibitem{Furlanetto 05}Furlanetto, S. R., Loeb, A. 2005, ApJ, 634, 1

\bibitem{Furlanetto 03}Furlanetto, S. R., Loeb, A. 2003, ApJ, 588, 18

\bibitem{Galama 98}Galama, T. J., Vreeswijk, P. M., van
Paradijs, J., et al. 1998, Nature, 395, 670

\bibitem{Greif 06}Greif, T., H., \& Bromm, v. 2006, MNRAS, 373, 128


\bibitem{Guetta 07}Guetta, D., \& Piran, T. 2007, JCAP, 07, 003


\bibitem{Haiman 00}Haiman, Z., Abel, T., Rees, M. J. 2000, ApJ, 534,
11


\bibitem{Heger 02}Heger, A., Woosley, S. E. 2002, ApJ, 567, 532

\bibitem{Hirschi}Hirschi, R. 2007, A\&A, 461, 571

\bibitem{Hjorth 03}Hjorth, J., Sollerman, J., Mller, P., et al.
2003, Nature, 423, 847

\bibitem{Hopkins 06}Hopkins, A. M.,  Beacom, J. F. 2006, ApJ, 651, 142

\bibitem{Hosokawa 11}Hosokawa, T., Omukai, K., Yoshida, N., Yorke, H.
W. 2011, Science, 334, 1250

\bibitem{Hosokawa 12}Hosokawa, T., Yoshida, N., Omukai, K., Yorke, H.
W. 2012, ApJL, 760, L37


\bibitem{Inoue 13}Inoue, Y., Tanaka, Y. T., Madejski, G. M. et al.
2013, arxiv: 1312. 6462


\bibitem{Ishida 11}Ishida, E. E. O., de Souza, R. S., \& Ferrara, A.
2011, MNRAS, 418, 500

\bibitem{Johnson 10}Johnson, J. L. 2010, MNRAS, 404, 1425

\bibitem{Kistler 08}Kistler, M. D., Y$\rm \ddot{u}$ksel, H., Beacom, J. F., Stanek, K. Z. 2008,
ApJL, 673, L119

\bibitem{Kistler 09}Kistler, M. D. et al. 2009, ApJL, 705, L104




\bibitem{Kulkarni 13}Kulkarni, G., Rollinde, E., Hennawi, J. F., Vangioni, E. 2013, arXiv:1301.4201

\bibitem{Lamb 00}Lamb, D. Q., \& Reichart, D. E. 2000, ApJ, 536, 1

\bibitem{Larsom 11}Larson, D., Dunkley, J., Hinshaw, G., Komatsu, E.,  Nolta, M. et al.
2011, Astrophys. J. Suppl., 192, 16

\bibitem{Mackey 03}Mackey, J., Bromm, V., \& Hernquist, L. 2003, ApJ, 586, 1


\bibitem{Madau 01}Madau, P., Rees, M. J. 2001, ApJL, 551, L27

\bibitem{Marassi 09}Marassi, S., Schneider, R. \& Ferrari V. 2009, MNRAS, 398, 293

\bibitem{Meszaros 10}M{\'e}sz{\'a}ros, P.
\& {Rees}, M.~J. 2010, ApJ, 715, 967


\bibitem{Mo 02}Mo, H. J. \& White, S. D. M. 2002, MNRAS, 336, 112


\bibitem{Murakami 05} Murakami, T., \& Yonetoku, D. 2005, ApJL, 625, L13


\bibitem{Ostriker 96}Ostriker, J. P., Gnedin, N. Y. 1996, ApJL, 472, L63


\bibitem{Porciani 01}Porciani, C., \& Madau, P. 2001, ApJ, 548, 522


\bibitem{PS 74}Press, W. H., Schechter, P. 1974, ApJ, 187, 425

\bibitem{Robertson 10}Robertson, B. E., Ellis, R. S., Dunlop, J. S., McLure, R. J., \&
Stark, D. P. 2010, Nature, 468, 49


\bibitem{Salvaterra 09}Salvaterra, R., Guidorzi, C., Campana, S.,
Chincarini, G., Tagliaferri, G. 2009, MNRAS, 396, 299

\bibitem{Scannapieco 01}Scannapieco, E., \& Broadhurst, T. 2001, ApJ, 549,
28

\bibitem{Springel 03}Springel, D. N., Hernquist, L. 2003, MNRAS,
339, 312

\bibitem{ST 99}Sheth, R. K., Tormen, G. 1999, MNRAS, 308, 119

\bibitem{Stacy 10}Stacy, A., Greif, T. H., \& Bromm, V. 2012, MNRAS,
422, 290

\bibitem{Stanek 03}Stanek, K. Z., Matheson, T., Garnavich, P. M., et
al. 2003, ApJL, 591, L17

\bibitem{Suwa 2011}
Suwa, Y. \& Ioka, K. 2011, ApJ, 726, 107

\bibitem{Tan 13}Tan, W. W., Cao, X. F., \& Yu, Y. W. 2013, ApJL, 772, L8

\bibitem{Tornatore 07} Tornatore, L., Ferrara, T., \& Schneider, R.
2007, MNRAS, 382, 945

\bibitem{Totani 97}Totani, T. 1997, ApJL, 486, L71

\bibitem{Trenti 09}Trenti, M. \&
Stiavelli, M. 2009, ApJ, 694, 879

\bibitem{} van Eymeren, J., Bomans, D. J., Weis, K. \& Dettmar R. J.
2007, A\&A, 474, 67


\bibitem{}Wang, F. Y. \& Dai, Z. G. 2009, MNRAS, 400, L10

\bibitem{}Wang, F. Y. \& Dai, Z. G. 2011, ApJ, 727, L34

\bibitem{Wang 13}Wang, F. Y. 2013, A \& A, 556, A90

\bibitem{Weinmann 05}Weinmann, S. M., \& Lilly, S. J. 2005, ApJ, 624, 526

\bibitem{Wijers 98}Wijers, R. A. M., Bloom, J. S., Bagla, J. S., \& Natarajan, P. 1998,
MNRAS, 294, L13






\bibitem{Yu 12}Yu, Y. W., Cheng, K. S., Chu, M. C., \& Yeung, S. 2012,
JCAP, 07, 023

\bibitem{Yuksel 08}Y{\"{u}}ksel,H., Kistler, M. D., Beacom, J. F., \& Hopkins, A. M. 2008, ApJ, 683,
L5


\end{thebibliography}
\end{document}